# Wireless communication, identification and sensing technologies enabling integrated logistics: a study in the harbor environment

**Mario G. C. A. Cimino**
University of Pisa, Department of Information Engineering, largo L. Lazzarino 1, 56122, Pisa, Italy

**Nedo Celandroni, Erina Ferro, Davide La Rosa, Filippo Palumbo**
National Research Council, Institute of Information Science and Technologies, via G. Moruzzi 1, 56124, Pisa, Italy

**Gigliola Vaglini**
University of Pisa, Department of Information Engineering, largo L. Lazzarino 1, 56122, Pisa, Italy

**ABSTRACT**

In the last decade, integrated logistics has become an important challenge in the development of wireless communication, identification and sensing technology, due to the growing complexity of logistics processes and the increasing demand for adapting systems to new requirements. The advancement of wireless technology provides a wide range of options for the maritime container terminals. Electronic devices employed in container terminals reduce the manual effort, facilitating timely information flow and enhancing control and quality of service and decision made. In this paper, we examine the technology that can be used to support integration in harbor's logistics. In the literature, most systems have been developed to address specific needs of particular harbors, but a systematic study is missing. The purpose is to provide an overview to the reader about which technology of integrated logistics can be implemented and what remains to be addressed in the future.

**Keywords:** *Wireless Communication, Harbor's Logistics, Marine Container Terminal, Identification Technology, Sensing Technology*

## 1. INTRODUCTION

In each country, the logistics and freight transport are key factors of competitive advantage for firms; today, they are undergoing significant transformation due to the innovation in the Information and Communication Technologies (ICT). These developments have made possible the emergence of a new model of logistic systems based on the concept of system integration. In essence, with the application of large-scale information systems, new models of organization, planning and management have been developed, integrating the various phases of the supply-production-distribution chain. In this way, the freight has joined - as a subsystem - the broader system of logistics. The modern port, which has been developed since 1980, is characterized by the containers traffic and its integration in a logistics network where land and sea segments are integrated. The efficiency of a port is strongly influenced by its ability to forge links with the hinterland to make the goods quickly converge towards their final destination. The computerization of the logistic and inter-port port processes is one of the key factors to support the improvement, the speeding, the safety and the streamlining of the information flows along the transport logistics chain, while encouraging the growing integration of the maritime transport with the other transport modes. The computerization of the ports is part of the initiatives of the European Union for an ever-expanding telematics in all modes of transport. The essential objectives of this computerization policy of the documents' flow between public (Customs and Finance Guard) and private (customs brokers, freight forwarders gap, container terminal, warehouse, truck drivers) entities involved in the process are: (i) reduction of manual steps in the processing of documents; (ii) faster and more secure exchange of information among the various actors; (iii) optimization of the workflow; (iv) compliance with the laws; (v) reduction in the cost and time management; (vi) minimization of the environmental impact on the different territorial systems; (vii) simplicity and fluidity, even administrative, of the goods' paths through the various infrastructures, studying different technological solutions; (viii) reduction in the road traffic congestion and maximum efficiency of multi-modality by the rational use of all the real-time information on the movements of goods and people; (ix) warranty of the transport safety.

A very important aspect in the logistic chain is the real-time tracking of goods (containers in the first place). In this context, the technology offers very advanced solutions, now consolidated and with sustainable costs. Wireless communication technologies, such as Global Positioning Systems (GPS), General Packet Radio Service (GPRS), and Geographic Information Systems (GIS), along with advanced Internet solutions, provide transparency and more specific information for the instant location and tracking of the shipments and their delivery status. Another fundamental element concerns the integrity checking of the goods (primarily, the container) and their not-burglary. Radio-Frequency IDentification (RFID) seals can be used to store and transfer the necessary information. In addition, the usage of wireless networks in large spaces and their integration with other systems (such as web services) allows provisioning such information in real time with respect to third parties (customs, financial police, operational partners, customers, etc.). An example of a location service and real-time tracking of goods is based on intermodality of freight: railway, road, and ship transport. Especially for dangerous goods, it is appropriate to provide an alarm in the event of accident and/or spillage of material and consequent alert of the appropriate Entities, giving them the georeferencing of the point where the accident happened, the communication of the type of material transported, and the identification of the nearest place with neutralizing agents available. Different technologies should be applied in a multimodal logistics environment [41], as exemplified in what follows: (i) an active GPS applied on the terrestrial load unit (wagon, container, or tank), which detects and constantly transmits its ground position to a service center; (ii) sensors controlling the products' status (thermometers, barometers, accelerometers and other sensors necessary for the real-time monitoring of the state of the goods); these sensors are installed on individual goods; (iii) RFID sensors for automatic switching to the customs gates and the reading of the documentation; (iv) a terrestrial Wireless communication system (Wi-Fi, WiMax) or a satellite system, to transmit data from the sensor boards with the relevant geo-referencing information to a service center; the service center collects the data and enrich them with other data, such as traffic information, and submit them selectively to the Operations Rooms of each body in charge; (v) alternative routes can be suggested to the terrestrial media in case of accidents, traffic jams, blocked roads, etc.; (vi) a communication satellite, preferably a geostationary satellite, to have guaranteed coverage transmission, once the goods are loaded onto ships.

The advancement of wireless technology provides a wide range of options for the maritime container terminals. In this paper, we examine the technology that can be used to support harbors logistics, with the purpose of providing an overview to the reader about which technology can be considered in integrated logistics of maritime container terminals.

The technologies presented in this paper are organized as follows. Section 2 is focused on Short-range wireless and sensing technology: Wireless Sensors Networks (WSN), Bluetooth, Wi-Fi, ZigBee. Long-range wireless technology is covered in Section 3: Cellular Networks (GSM, UMTS, and LTE), Worldwide Interoperability for Microwave Access (WiMAX), and Terrestrial Trunked Radio (TETRA). Section 4 is devoted to Satellite Wireless Technology: Global Positioning System (GPS) and Digital Video Broadcasting-Return Channel via Satellite (DVB-RCS). Finally, Wireless Identification and Sensing Technology is studied in Section 5: Radio Frequency Identification (RFID) and Remotely Piloted Aircraft (RPA).

## 2. SHORT-RANGE WIRELESS AND SENSING TECHNOLOGY

### 2.1 Wireless Sensor Networks

Wireless Sensor Networks (WSNs) are characterized by a distributed architecture, where a set of autonomous electronic devices is able to fetch data from the surrounding environment and to communicate among them. Sensor networks are a natural, yet revolutionary evolution of the use of sensors in industrial environments. The market, in fact, requires devices and systems with increasingly higher capacities and high levels of functionality. The sensors used in these devices and systems are typically used to estimate a physical quantity or used to monitor parameters of "process control". The use of a network of transducers produces undeniable advantages compared to the use of traditional sensors in terms of flexibility, performance, ease of installation, costs of possible future developments, and maintenance activities. The need of implementing a network infrastructure at the same time, however, requires the use of more advanced sensors that are no longer simple transducers of physical quantities, but more complex systems that integrate, other than the measurement capability, also storage capacity, calculation capacity and, of course, communication interfaces. These observations lead to the definition of "smart sensor", which are integrated devices equipped with a micro-controller able to perform activities of communication and information processing.

The WSN are poorly invasive, given the limited size of the nodes and the absence of cabling and allow for quick installation in environments where complex network infrastructures are not feasible. The WSN can be programmed to be able to self-organize according to various network topologies (star, linear, cluster, mesh, etc.), selected according to the specific application context. A sensor network is a set of sensors disposed in the vicinity or inside of the phenomenon to be observed. These small devices are produced and distributed *en masse*, have a negligible cost of production and are characterized by very small size and weight. Each sensor has a limited and non-renewable reserve of energy and, once put in place, it must work independently; for this reason, these devices must constantly maintain as low as possible the power consumption, in order to have a greater life cycle. To obtain the greatest possible amount of data, it is necessary to carry out a massive distribution of sensors (in the thousands or tens of thousands) in order to have a high density (up to 20 nodes / $m^3$) and to ensure that nodes are all close the one with the other, a necessary condition to allow communication. The communication, carried out through short-range wireless technology, is usually asymmetric because the nodes send the collected information to one or more special nodes of the network, said sink nodes, which collect data and transmit them typically to a server or to a computer. The communication may take place autonomously by the node when a given event occurs, or may be induced by the sink node by sending a query to the nodes concerned. In Fig. 2.1 and Fig. 2.2 two examples of WSN setting are shown.

### 2.1.1 Classification of Wireless Sensor Networks

One of the typical distinctions that are made in the field of WSNs concerns the topology of the network. It is possible, indeed, to distinguish between *homogeneous* topology (where all nodes provide the same functionality) or *hierarchical* topology (where specialized nodes perform specific functions). A typical division of the tasks in a hierarchical WSN sees some nodes engaged in the task of monitoring, collecting data on the physical quantity of interest, and other nodes involved in the data processing and routing of results towards the base station. Although the hierarchical structure seems much more suited to the needs of a WSN, it has some inherent drawbacks. First of all, computing resources and the energy required could be significantly different depending on the location of the node in the hierarchy, thus creating the need to use different sensor nodes depending on the task assigned. In addition, the rigid division of tasks goes against the needs of network resiliency, as the tasks of any faulty node could not be taken over by another sensor node topologically close to him, but hierarchically different. A hybrid solution is that of the *clustered* structure. In this case, the task of each node depends on its spatial and topological location within the geographical area to be analyzed. Indeed, in many applications for WSNs, as in the case of the environmental monitoring, building automation or installations of industrial production (manufacturing plants), it is possible to identify the areas to be monitored geographically separates (e.g. trees, rooms, robot). In these cases, the sensor nodes are scattered within these objects to collect uniform data (e.g. temperature, vibration) and must report such data to a base station located in an arbitrary position, in all likelihood topologically remote.

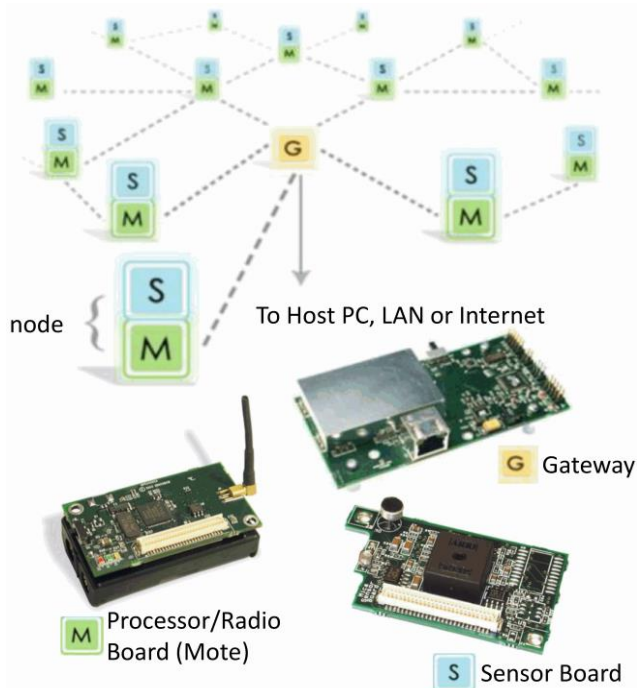

Fig 2.1: Typical Wireless Sensor Network's components (nodes): Gateway, Motes, and Sensor Boards.

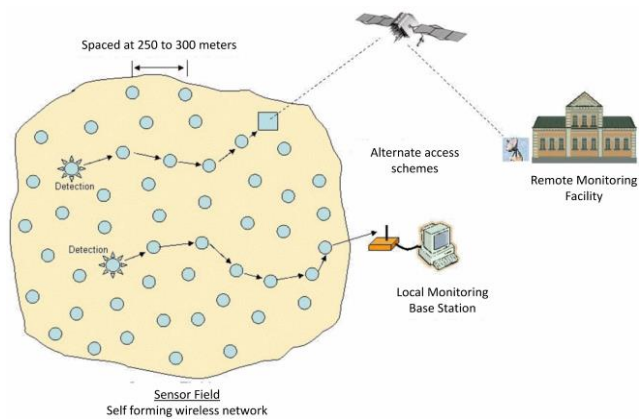

Fig 2.2: A typical deployment scenario for remote monitoring.

Another possible classification of WSNs is the potential mobility of the nodes. It is defined as *static* a network where nodes remain in their original geographic location (or, at least, topological), while *dynamic* is a WSN where nodes can be arbitrarily moved. Finally, another distinction, is between *centralized* and *distributed* WSNs. In the first case, the tasks will be allocated to particular nodes (for example, data processing could be completely left to the base station), in the second, it will be the network itself to provide, by distributing the tasks to all the nodes that compose it. Despite the fact that a centralized structure seems easier and more efficient, it is important to note that a sensor network is often composed of homogeneous devices, from the point of view of both hardware characteristics and Operating System, and therefore loading only some of the nodes of specific tasks would introduce strong differences in the structure of the network, thus compromising its longevity.

### 2.1.2 Metrics evaluation for wireless sensor networks

In order to objectively evaluate the goodness of a given system architecture for WSNs in its entirety (i.e. taking into account the hardware, software and middleware components that allow to achieve a given functionality) and to be able to compare with other solutions, it is necessary to define the criteria for the judgment.

Important metrics are: (i) *lifetime*, which expresses the total time of operation of an application without the need for any external intervention (e.g., to recharge or replace the batteries); (ii) *cost* and *deployment* needed for the creation of the network; (iii) difficulties in *managing* and *maintenance*; (iv) *size of the network*, because the ability to miniaturize the components is very important in this area; (v) *robustness*, the ability to tolerate critical situations such as, for example, the malfunction of some of the nodes, the interference on the communication channel and the reduction in the amount of available energy; (vi) *reactivity*, i.e., the ability to react quickly to external stimuli, both at the level of the node at the network level; (vii) *simplicity* of the project, use, creation, and management; (ix) *security*, i.e., forbid the control when not authorized; (x) *confidentiality* and *integrity*, i.e., to protect and keep data consistent; (xi) ability to function in a *distributed* manner; (xii) *routing* protocols.

References [1, 2, 3, 4, 5, 6] deep analyze some of aspects of performance evaluation in WSNs.

### 2.2 Bluetooth

Bluetooth is an industrial data transmission technology for WPAN (Wireless Personal Area Network) [7]. It provides a standard, economical and safe way to exchange information between different devices through a secure short-range radio frequency. Bluetooth (sometimes abbreviated as BT) searches for devices covered by the radio signal within a radius of a few tens of meters by placing them in communication each other. These devices can be e.g. PDAs, mobile phones, personal computers, microphones, laptops, printers, digital cameras, game consoles, etc., provided that they carry the specific hardware and software required by the standard. The Bluetooth specification has been designed with the primary goal of getting low power consumption, a short range (up to 100 meters of coverage for a Class 1 device and up to one meter for Class 3 devices) and a low-cost production for compatible devices. The Bluetooth protocol works in the free frequencies of 2.45 GHz. The version 1.1 and 1.2 of the Bluetooth handles transfer speeds of up to 723.1 kbit / s. Version 2.0 runs a high-speed mode that allows up to 3 Mbit/s. This mode, however, increases the power consumption. The new version (Bluetooth Low Energy or Bluetooth Smart) is able to halve the required power with respect to Bluetooth 1.2 (with the same traffic sent). Bluetooth is a standard comparable with Wi-Fi, since this is a protocol developed to provide high-speed transmission with a larger radius, at the cost of greater power dissipation and a lot more expensive hardware. In any case, the Bluetooth standard also includes long distance communications between devices to make wireless LANs. Each Bluetooth device is able to simultaneously manage communication with other 7 devices although, being a link-type master-slave, only one device at a time can communicate with the server.

### 2.3 Wi-Fi

Wi-Fi is a telecommunication technology that enables end users to connect with each other through a local network wirelessly (WLAN) based on IEEE 802.11 standard (standards.ieee.org/about/get/802/802.11.html). In turn, the so-obtained local network can be connected to the Internet via a router and can use all the connectivity services offered by an Internet Service Provider (ISP). Any device or user terminal (computer, mobile phone, PDA, tablet, etc.) can log on this network type when integrated with the technical specifications of the Wi-Fi protocol. The Wi-Fi network is a telecommunications network, possibly interconnected with the Internet, conceptually similar to a cellular network covering a small-scale (local), with rice-transmission radio devices such as access points (AP) in lieu of the traditional radio stations, base of mobile networks (client-server architecture model). To increase the connectivity range of a single access point (about 100 m), whose transmission power is limited by specific regulations related to the health electromagnetic risk (100 mW), and thus to be able to cover the desired area, several Access Points (and related cell coverage) are cabled together in the local network. The radio part or the radio access point-user interface is the access network and the wired LAN that connects all the access points represents the transportation network. The cells of coverage of the AP are often partially overlapped in order to avoid coverage holes of the signal by creating a total coverage area called ESS (Extended Service Set), while the cabled part is generally a wired Ethernet network that can be of shared bus type or switched. Thus, the global network obtained can be connected to the Internet through a router, taking advantage of its internetworking services. Architectural solutions are also possible without a wired backbone that directly connects in wireless mode the access points thus enabling their communication as a wireless distributed system, i.e. with exchange of information entirely through radio interfaces, albeit with a loss in spectral efficiency of the system. Completely wireless architectures with no access points (peer-to-peer architecture model) are also possible, with each base station that receives / transmits directly to or from other stations (Independent Basic Service Set, or IBSS ad-hoc mobile). Architectural wireless solutions obviously entail costs and construction times considerably low, at a price of a lower connection performance. The difference of the Wi-Fi networks with respect to other cellular coverage networks resides in the communication protocols, namely the protocol stack that redefines the first two levels (physical and link), or the physical layer protocols and the multiple access protocols to the radio medium. In particular, since the transmission of each station occurs at the same operating frequency (2.4 or 5 GHz), to avoid collisions in the reception, the protocol of multiple access CSMA / CA (Carrier Sense Multiple Access with Collision Avoidance) is used~\cite{othman2005multi}. Ports and airports are great places to provide free Wi-Fi [9, 10]. As an example we cite the port of Trieste, whose infrastructure is made up of a modern fiber-optic network, a wireless backbone, and a WI-FI network open to users and operators.

### 2.4 ZigBee

The IEEE 802.15.4 ZigBee standard has been defined to meet the needs of the market with regard to a flexible wireless network technology, standardized and low-cost, able to offer a reduced power dissipation, good reliability, security, and interoperability in control and low bit-rate monitoring applications. The standard defines the protocol and the interconnection of devices via radio communication in a Wireless Personal Area Network (WPAN). The main features of a WPAN are ease of installation, reliability of data transferred, the short-range operation, the low cost, the reasonable battery life, and finally a simple and flexible protocol. The ZigBee protocol is defined by a layered architecture based on the ISO-OSI model; each layer is responsible for a part of the standard and provides services to the upper layers. The standard defines the specification of the physical layer and the access control to the medium. The top layers are treated by the organization Zigbee Alliance (www.zigbee.org/zigbeealliance), a consortium of companies that work together with no profit in order to promote the development of an open global standard for the wireless control of devices to be used in home and industrial automation.

Depending on the application requirements, an IEEE 802.15.4 WPAN can operate in two topologies: star and peer-to-peer. In the star topology, communication is established between devices and a single control center, called the PAN Coordinator. Communication occurs only between the coordinator and full-function devices (FFD) or reduced-function devices (RFD). The star network is independent of, the other star networks that operate at that time, and this result is obtained by choosing as PAN coordinator one that is not already currently used in another PAN, within the radio influence range. The PAN coordinator allows the addition of other devices to the network, either FFD or RFD. Even in the peer-to-peer topology there is a coordinator but it differs from the star topology because, in this case, any device can communicate with any other device, provided that they are in the same range. This topology allows the formation of more complex networks, such as the mesh networks. A peer-to-peer network can be ad hoc, self-manageable, autonomous and reconfiguring. It may also allow a multi-hop network to route messages from any device to any other device on the network.

Data transfer can be done in three ways: 1) to transfer from one device to the coordinator; 2) to transfer from the coordinator to a device that receives them; 3) data transfer between two devices. In a star topology, only two types of transfer are feasible because, in this case, the communication is carried out only between the device and the coordinator; in a peer-to-peer topology all three transactions are possible.

The physical layer is a key part of a ZigBee device. The ZigBee devices operate in the 2.4 GHz ISM bands in the world and in the 915 MHz and 868 MHz ISM bands in America and Europe, respectively. In the IEEE 802.15.4 standard (year 2006) the 868/915 MHz optional channel band was introduced, with ASK or O-QPSK modulation. To support the growing number of channels, the assignment must be defined through a combination of channel numbers and "pages" of channels.

ZigBee 3.0 is the unification of the Alliance's market-leading wireless standards into a single standard. This standard will provide seamless interoperability among the widest range of smart devices and give consumers and businesses access to innovative products and services that will work together seamlessly to enhance everyday life.

ZigBee 3.0 is currently undergoing testing and is expected to be available in Q4 of 2015. ZigBee 3.0 simplifies the choice for developers creating Internet of Things products and services. It delivers all the features of ZigBee while unifying the ZigBee application standards found in tens of millions of devices delivering benefits to consumers today. ZigBee 3.0 standard enables communication and interoperability among devices for smart homes, connected lighting, and other markets so more diverse, fully interoperable solutions can be delivered by product developers and service providers. ZigBee 3.0 is based on IEEE 802.15.4, which operates at 2.4 GHz (a frequency available for use around the world) and uses ZigBee PRO networking to enable reliable communication in the smallest, lowest-power devices. ZigBee 3.0 defines more than 130 devices and the widest range of devices types including home automation, lighting, energy management, smart appliance, security, sensors, and health care monitoring products. It supports both easy-to-use DIY installations as well as professionally installed systems. All current device types, commands, and functionality defined in current ZigBee PRO-based standards are available in the ZigBee 3.0.

Some aspects of ZigBee are deeply treated in [11, 12, 13]; examples of its application are in [14, 15, 16].

## 3. LONG-RANGE WIRELESS TECHNOLOGY

### 3.1 Cellular Networks: GSM, UMTS, and LTE

In telecommunications, GSM (Global System for Mobile Communications, originally "Groupe Spécial Mobile"), is the 2G (2nd generation) standard for mobile phone and currently the most widespread in the world: more than 3 billion people in 200 countries use GSM mobile phones via the namesake cellular network. In particular, GSM is an open standard, developed by CEPT and ETSI, designed and maintained by the 3GPP consortium (including the ETSI part). The introduction of GSM has been a real revolution in the field of cellular telephone systems. Basically, the numerous advantages compared to previous cellular systems were: (i) interoperability among different networks that are part of a single international standard; (ii) digital communication. The introduction of a digital type transmission has three important consequences: (i) higher transmission speeds, thanks to data compression techniques typical of source coding (LPC encoding); (ii) new wider services (e.g. SMS) by increasing the transmission speed; (iii) safety functions in terms of encrypting communications. Additional information on GSM can be found in [44, 45].

The Universal Mobile Telecommunications System (UMTS) is the digital cellular mobile radio system of the third generation, evolution of GSM [39]. Conceived as a global system comprising both terrestrial and satellite components, it is capable of supporting a data transmission rate of 2 Mbit /s, thus making possible a series of multimedia services, such as the ability to send faxes and e-mail, the Internet access, the download and the transmission of data packets without the need of a fixed terminal, the video conferencing, and to use of the video telephony. UMTS, unlike GSM, which for the data transmission uses circuit-switched technology, integrates circuit and packet data transmission; this allows obtaining different services, such as continuous virtual connections to the network and alternative modes of payment (for example payments proportional to the number of bits transferred or to the used bandwidth width). UMTS is sometimes launched on the market with the designation 3GSM to highlight the combination between the 3G technology and the GSM standard, which should in the future be completely replaced by 3GSM. Compared to the previous GSM system, the improvement consists in the increased transmission speed due, in turn, to both the adoption of a multiple channel access of W-CDMA (Wideband-Code Division Multiple Access) type, more efficient than the TDMA of GSM from the spectral efficiency view point, and the use of more efficient modulation numeric schemes. By using W-CDMA, UMTS supports a theoretical maximum transfer rate of 21 Mbps, although users of the existing networks have an available transfer rate up to 384 kbps, and up to 7.2 Mbit / s using R99 devices and HSDPA devices (in download connections), respectively [40].

LTE (Long Term Evolution) indicates the most recent evolution of mobile telephony standard GSM / UMTS, CDMA2000 and TD-SCDMA. Born as a new generation system for broadband mobile access from the theoretical point of view it is part of the Pre-4G segment, standing in an intermediate position between the 3G technologies, such as UMTS, and those of the pure fourth generation (4G). Nevertheless, the ITU (International Telecommunication Union) has recently decided to apply the term 4G also to LTE. HSPA (High Speed Access Packet, a family of protocols for mobile telephony that extend and enhance the performance of UMTS) and LTE are strong antagonists of WiMAX and its evolution; the availability of large-scale HSPA and LTE in major urban areas has reduced the prospects of success of WiMAX, especially as applied in the field of Internet and mobile broadband. Many CDMA2000 operators are planning to switch to the LTE standard as soon as the devices become available, thus abandoning the CDMA, whose success is now more and more limited, making it much closer to the possibility of creating a truly global standard for mobile communications. Some aspects of LTE are treated in references [35, 36, 37, 38].

## 3.2 Worldwide Interoperability for Microwave Access

In telecommunications, WiMAX (Worldwide Interoperability for Microwave Access) is a technology and a technical standard that allows the wireless access to broadband telecommunications networks.

The acronym was defined by the WiMAX Forum (formed in June 2001), a consortium of more than 420 companies whose purpose is to develop, oversee, promote and test the interoperability of systems based on the IEEE 802.16 standard, also known as Wireless MAN (Wireless Metropolitan Area Network).

WiMAX is a broadband wireless transmission technology. Like other wireless technologies, it can be used on many types of landscapes (the connection is set by the provider). WiMAX is based on the IEEE 802.16 standard family, also known as WirelessMAN, specialized in broadband wireless access of point-to-multipoint type [42, 43].

Depending on the legislation of the country of reference, the frequencies used by WiMAX may be subject to licensing (in Italy it was the connection used by the Navy). The WiMAX technology does not necessarily require optical visibility but, without it, the benefits are much lower and connectivity is restricted to limited areas. On the basis of expectations on WiMAX, it was expected to provide broadband coverage for a wide range (up to 50km) from each base station, with the consequent possibility of using technology to reduce the digital divide. The field tests showed that on the 3.5 GHz frequency, in optical visibility conditions, the performance is acceptable for maximum distances of a few kilometers, which is reduced to a few hundred meters in conditions of zero sight visibility. There are also complications due to electromagnetic legislation, which varies from state to state, which could limit the exploitation of WiMAX for such purposes. WiMAX 2 (WirelessMAN-Advanced or WiMAX-2) is the IEEE 802.16m standard; it represents the second generation of WiMAX technology that should allow reaching the threshold of 300 Mbps downstream (4G, fourth generation technology).

## 3.3 Terrestrial Trunked Radio

TETRA (Terrestrial Trunked Radio, originally Trans European Trunked Radio) is a set of standards for private telecommunication systems (Professional Mobile Radio, PMR), addressed to professional users (the public security forces, fire brigades, civil protection), but also to service providers (transport, energy) interested in having their own mobile network. TETRA is an ETSI standard, whose first version was published in 1995, recommended by the European Radio Communications Committee (ERC).

The TETRA networks provide the typical services of the private networks: voice calls of half-duplex group with Push-to-talk (PTT), dynamic management of the groups to which they belong, queuing and pre-emption of the calls according to priority, call authorized by dispatcher, environment listening, status messages, localization via GPS. Other features include the typical services of the cellular networks: individual full-duplex calls, caller identification, short text messages (SDS), redirection of the busy or unreachable user. The TETRA terminals can also act as cellular phones if they reach an external PSTN, ISDN or PBX network via gateways. In addition to voice communications, data communications are possible in circuit-switched or packet-switched, but still at low transmission speed (never up to 28.8 Kbps gross, though using a whole carrier). Privacy or confidentiality of communications is achieved by transmissions encryption in the air using a unique key common to all users, or individual and group keys regenerated on session basis. The end-to-end user encryption is also supported. An updated version of the ETSI standard TMO, called TETRA 2, was released, which adds support for airborne radio and especially specifies a radio interface using the OFDM technology, an adaptive QAM modulation, and bandwidths up to 150kHz, thus allowing to reach data rate speeds up to 538 Kbps.

The main advantages of TETRA over other cellular technologies (such as GSM) are: (i) very high levels of coverage with a small number of transmitters and consequent reduction in infrastructure costs, thanks to the use of a lower frequency; (ii) quick connection set-up time: a one-to-many call is usually created in 0.5 seconds (typically less than 250 ms for a single node) against the many seconds required for GSM; (iii) different ways of emergency with respect to other cellular technologies, with the ability for a base station to process local calls in the absence of the rest of the network and the Direct Mode feature, where the terminals can continue to directly share the channel if the infrastructure has a failure or is unreachable; (iv) the gateway mode, in which a single terminal connected to the network can act as a relay for the other neighbor terminals that are unable to get in touch with the infrastructure; (v) the point-to-point functionality, which other radio services do not offer for emergencies, thus allowing the user a link between terminals without the direct involvement of a supervisor operator or a dispatcher; (vi) the one-to-one connections together with the one-to-many and many-to-many connections, unlike other cellular technologies, which only allow one-to-one connections; these multiple operating modes are important for public security and professional purposes.

The main disadvantages of TETRA with respect to the other cellular technologies are: (i) it can support a relatively low density of users per area, compared to GSM and other technologies (although this is not usually a real limitation for professional applications it is addressed to); (ii) terminals are more expensive due to the absence of a real market request (given the limited number of users) and for the high level of quality required for high-reliability systems; (iii) the data transfer rate is only 7.2 Kbps per timeslot (3.5 Kbps net throughput of data packets), although up to 4 timeslots can be combined into a single data channel to achieve higher speeds; this is due to the fact that the constraint imposed by the spectrum spaced at 25 KHz must be respected; (iv) due to the impulsive nature of the TDMA protocol used, the terminals can interfere with other electronic devices if they are very close, such as pacemakers, defibrillators and other devices, radio transmitters (generally less than one meter distant).

## 4. SATELLITE WIRELESS TECHNOLOGY

### 4.1 Global Positioning System

GPS (Global Positioning System) is a civil system of satellite navigation and positioning. Through a dedicated network of artificial satellites in orbit, it provides a mobile terminal or a GPS receiver with the information on the current geographic coordinates and time, provided that there is an unobstructed contact with at least four satellites [17]. The localization is done through the transmission of a radio signal from each satellite and the processing of the signals received by the receiver. In order to always have a very high precision, the receiver is periodically synchronized with the atomic clocks on the satellites in view.

The use of GPS has brought several benefits to the transportation process including: (i) constant tracking of the vehicles' position; (ii) fast communication between drivers and headquarters; (iii) increased speed of the transport process; (iv) easy electronic archiving of the documents; (v) increased transparency, timeliness, and accuracy of information.

The GPS Tracker (or satellite tracker) is a system that uses the GPS signal to identify the exact location of a vehicle, person or any other object it is associated to [18]. The operation takes place thanks to the network of GPS satellites that constantly send signals and, according to their intensity, a GPS receiver is able to identify the coordinates that determine the position. The locator is also equipped with a telephone circuit to connect to the GSM network through which it is possible to send commands to the device and requests for information, so it is necessary that is inserted into an ordinary GSM Global System for Mobile communications) phone card. The location data can be sent via a simple text message using the GSM / GPRS (General Packet Radio Service) leading provider of mobile network. This way, the position of the GPS tracker may at any time be displayed on a cartographic map such as Google Maps (for monitoring in real time). The text message can be sent even in conditions of poor GSM coverage and discontinuous. When needed, the data flow can however rely on the GPRS connection as a continuous data stream.

The GPS Tracker is typically used to trace the paths followed by trucks and trains carrying containers, but could also be used for monitoring the containers into the sea. This device offers several advantages: (i) reception of a single location by making a simple phone call to the GPS Tracker, which will respond via SMS sending all the information needed to establish its correct localization, latitude, longitude, speed, date, and time; (ii) reception of an SMS from the GPS tracker at regular time intervals (for example, get the location every 5 minutes); (iii) reception of an SMS when a threshold speed is exceeded (e.g. lorries transporting special materials need not to exceed specific and restrictive speed limits; (iv) reception of an alarm if the GPS tracker exits outside of a predetermined "region", that is, in the case where the shipment begins to move in a different direction with respect to the intended path or it is stopped for a period of time longer than the target; (v) reception of information about the battery level with the possibility to set/unset the sleep mode (energy saving); (vi) reception of an alarm triggered by specific events, e.g., when the container is opened or if there is an illegal intrusion in it (even when the doors are closed); (vii) configuration of the parameters APN (Access Point Name), username, password, etc. to get the localization information via GPRS;

The GPS system was initially designed to monitor in real time the position of the "fleet" [19]. Today the GPS is highly reliable and accurate, ensuring by means of the satellite signals the exact position of the vehicle where the receiver has been installed, with a position error of less than 1-2 meters. The locator is designed to offer maximum reliability even in extreme conditions, such as high temperature variations, high humidity, moisture, and vibrations. The continuous technological evolution has allowed, over the years, the realization of devices with increasingly reduced consumptions and especially able to manage the sleep-mode and switched back in time functions, indispensable on vehicles not equipped with their own power supply. In the case of trailers and containers, devices placed in suitable sealed containers with back up batteries are used. The antennas can be internal or external. The power supply for recharging the internal battery may be carried by the semi-trailer services or by photo-voltaic cells. The service allows detecting the attachment of the trailer to authorized vehicles and finding the means if removed from the tractor. The change in computation time when GPS is switched-on depends on the operating conditions at that moment (surrounding environmental obstacles and receivable satellites at that time and at that point).

A GPS receiver must "see" at least 4 satellites to calculate the position, but other factors come into play; as an example, many GPS receivers keep in memory the last recorded location at the switching off time to speed up the computation when they restart. If the GPS is turned off and transported away from the area where it calculated the stored position, this may prolong the updating phase. This is because the system starts processing the data starting from the last known element, which turns out to be different. Other reasons may be related to the movement of the receiver: the position calculation is considerably prolonged if the GPS is turned on in a moving vehicle, or a few moments before departure; even devices of excellent quality may exceed 10-15 minutes. Finally, the increasing ether congestion, saturated with radio transmissions, must be mentioned. A source of near and powerful radio-frequencies in some cases may disturb a GPS receiver, due to both the radio spectrum saturation and the level of electromagnetic interference on the part of the electronic equipment.

The data received from the GPS system are: (i) latitude in degrees, minutes; (ii) longitude in degrees, minutes; (iii) elevation in feet; (iv) current speed; (v) satellites visible from the GPS. To achieve a complete localization system, usually software with integrated cartography is used; this software is available on the computers at the headquarters, and allows following the fleet on the map and controlling all the functions of the remote locator (Fig. 4.1). In general, the GPS receiver can also be associated to other instrumentation, such as wireless devices (tags) for the recognition of containers connected to the tractor. The tag is equipped with accelerometer and activates during the movement of the container; it transmits its identifier together with sensory information, to the receiving system present in the tracker. Environmental obstacles are a first problem associated with GPS, both in the immediate vicinity and relatively more distant. Metal, concrete, rock and soil constitute an insurmountable obstacle to the flow of the GPS radio waves. Vegetation usually does not constitute a major obstacle provided it is not too thick and a high rate of moisture on the foliage is not present. The reception may be difficult in a mountainous area, where high cliffs may surround the point of reception. Another limitation is due to the satellite radio waves reflected against obstacles. A very important issue relates to the cost; in fact using GPS devices for each container is normally very expensive.

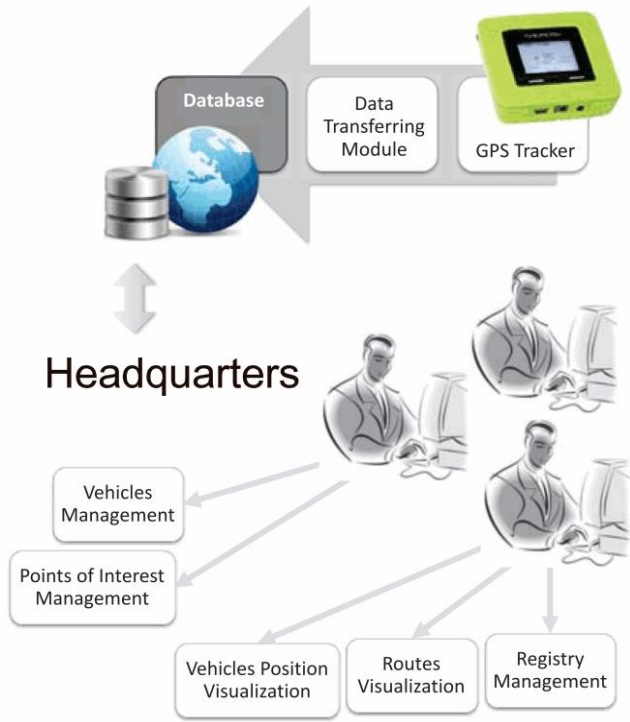

**Fig 4.1: A general scheme of a GPS Tracker based monitoring system.**

### 4.2 DVB-S/S2/RCS+M

For the purpose of better organizing the supply chain, it is important to have real-time knowledge of the state of the transport activities that can be carried out through the integrated use of satellite platforms for telecommunications and positioning, and the subsequent dissemination of data to all entities concerned over the Internet. There are many different systems that combine location with telecommunications; these systems are based, generally, on a satellite platform for the location of vehicles on the road and on a terrestrial infrastructure (typically the cellular network GSM / GPRS) for receiving and transmitting data from operational centers. The decision to adopt a satellite infrastructure as a means of communication rather than a mobile phone network (typical navigation and communication systems still on the market) is due to multiple factors: (i) a satellite system offers, with continuity, a wide coverage area (including the whole of Europe, in geostationary), thus covering even those areas for which it is not possible to implement a land cover; (ii) a satellite system provides a high-bandwidth, thanks to which the transfer of data and images, also of enormous size, occur in a very short time; (iii) it is possible to have a continuous contact with the service center, operated by the companies involved in the production chain, and the fleet of vehicles, which is not possible by using the cellular mobile radio system as a means to support communications for the geo-localization, as the phone networks may be congested or have holes in coverage.

Once decided to adopt a system of satellite coverage, it must be decided whether to use a geostationary satellite or a Leo (Low Earth Orbit) constellation, whereas the users mainly consist of moving vehicles, and therefore the type of services will be that of the multimedia messaging system (MMS).

The main actors are the means in motion that will have access to the DVB-RCS (Digital Video Broadcasting-Return Channel via Satellite) satellite system. DVB-RCS is a specification for a system of interactive communication satellite on-demand media formulated in 1999 by the DVB consortium. The first activities of the consortium were devoted to the development of technical specifications for the implementation of digital television by satellite DVB-S, which is currently used by most satellite operators in the world. On the tenth anniversary of the birth of DVB, DVB-S2 was born, second-generation system for broadcasting via satellite [29].

The system DVB-S2 [30] is designed for various applications satellite broadband: broadcasting services of standard definition TV (SDTV Standard Definition TeleVision) and high definition (HDTV High Definition TeleVision), interactive applications for residential and professional users, including access to the Internet, professional services of TV contribution and SNG (Satellite News Gathering), distribution of TV signals to digital terrestrial transmitters VHF / UHF, distribution of data and Internet sites (Internet trunking). There are three key concepts according to which the standard DVB-S2 has been defined: (i) increased transmission capacity compared to first-generation systems and in particular to the DVB-S; (ii) total flexibility; (iii) reasonable receiver complexity. To get the balance between performance and complexity, DVB-S2 uses the most recent developments in channel coding and in the modulation. The +M version added several new features, like the ability to use burst in DVB-S2 link back to the satellite, which includes techniques to mitigate the degradation of signal transmission and other solutions to combat signal loss on short rays. In March 2014 the DVB-S2 specification became a 2-part document to accommodate the new DVB-S2 Extensions (DVB-S2X) [31]. The DVB-S2 will remain a stand-alone specification and the DVB-S2X specification is optional.

## 5. WIRELESS IDENTIFICATION AND SENSING TECHNOLOGY

### 5.1 Radio Frequency IDentification

RFID (Radio Frequency IDentification) is a technology for the identification and the automatic data storage of objects, animals or people (AIDC Identifying and Automatic Data Capture), based on the ability of storing data by special electronic devices (called tags or transponders) and the ability of these devices to answer the query at a distance, by appropriate fixed or portable equipment communicating the information therein contained. Therefore, they can be assimilated to wireless "reading/writing" systems, with numerous wireless applications.

The RFID tag can be active, passive, semi-passive or semi-active. If *active*, it has: (i) a battery to power it; (ii) one or more antennas to send the signal to read and to receive responses even on different frequencies; (iii) one or more transponder / RFID tags and it can contain sensors; (iv) generally, a greater operating distance than the passive tag, arriving at a maximum of 200 meters.

If the RFID is *passive*, it simply contains a microchip (with unique identifier and possible memory), free of electrical power, an antenna and a material, called "substrate", that acts as a physical support and that is fed at the passage of a reader that emits a signal radio at low or medium or some Gigahertz frequencies. The radio-frequency activates the microchip and provides the energy needed to respond to the reader, back transmitting a signal containing information stored in the chip. If it is *semi-passive*, it is equipped with battery only used to power the microchip or auxiliary equipment (sensors) but not to power a transmitter, as in transmission a passive tag behaves as a passive RFID label. If the RFID is *semi-active*, it has a battery that powers the chip and the transmitter where, in order to save energy, the RFID tag is disabled, and it is enabled through a receiver with passive-tag technology; therefore, in the absence of queries, the tag can operate for long times.

The main element that characterizes an RFID system is the tag. The antenna receives a signal that it turns it into electricity, which powers the microchip. The so-activated chip transmits through the antenna (signal transmission circuit) the data therein contained to the apparatus that receives the data. In summary, an RFID tag is able to receive and transmit via radio frequency information contained in the chip to a RFID transceiver. The reader emits an electromagnetic/electric field that generates a current that powers the chip, through the process of induction in the tag. The so powered chip communicates all its information by means of the antenna to the reader, and the reader can also write data to the tag.

RFID technology has some advantages over traditional simple technology of bar-codes and magnetic stripes: (i) it must not be in contact to be read as the magnetic stripes; (ii) it is not necessary to be visible to be read as the bar-codes; (iii) information on the chip can also be added depending on the type of chip: read Only (information can only be read), write once-read many (the information in the chip can be written once but read an unlimited number of times, wead and write (information can be read and stored for a large but limited number of times); (iv) identification and verification occur in 1/10 of a second; (v) communication can be clear or encrypted.

In the harbor scenarios, RFID technology is used for the identification code of the containers; in particular, it provides the use of an electronic seal to be applied on each container unloaded from a vessel [20]. The studied RFID scenario involves the use of electronic seals, which are active tags (i.e. equipped with a battery for a greater operating range), able to identify the container and to record any tampering to the load. For the full functionality of the application, all operators and the mobile media must be equipped with portable RFID readers, while all accesses to the terminal deposits and the entry and exit gates must be equipped with specific reading gates. For the management of the containers, two technological scenarios can be considered. The first scenario involves the use of active RFID tags with the seal function, reusable, affixed to the container at the landing time and drawn at the customs clearance time. The second scenario presents the integration of the supply chain, in which each container is assumed to arrive at the port already equipped with an electronic seal installed upstream of the supply chain (by the shipping company or the person who sends the goods) and, consequently, there is a need to pick it up at the customs clearance time.

In the first scenario, the device is affixed by the custom agent who, near the pier, makes the so-called operation "seen landed" and it is taken when the goods, after the customs clearance, permanently leaves the terminal. At the same time, the antennas, placed in strategic points of the port, allow precisely tracking all the movements of the tagged containers and updating in real time the data in the Customs Agency information system, in line with the terminal operator. The data managed through the RFID tags make clear and transparent all the goods movements, and, above all, they are constantly tracked, which means that information and interventions are updated from time to time, recorded and dated. Reading the tag, it is possible at any time to go back to any type of data necessary to understand the origin and the provenance of the products, but also the type of contents of packages, in turn stored in the pallet, packed in warehouses and then taken into care for the resolution of the orders and their delivery.

### 5.1.1 Benefits of the RFID technology in the port

Using RFID technology, the possible benefits include: (i) the increase in productivity in the identification of the container code; (ii) the verification of the seal integrity at the port entry and exit gates; (iii) the automation of the activities and the reduction in human errors; (iv) the reduction in the ship loading and unloading time; (v) the electronic seal, being hardly alterable and recording any violations, to deter unauthorized opening of the load.

By adopting the RFID solution, every single container can be monitored at all stages of the journey, from the moment it is loaded on the ship until the arrival at the destination port. Therefore, each container will be equipped with an RFID tag, which will broadcast regular information about the load location, condition, and integrity. This information is picked up by means of a GSM terminal installed on board the ship and from there, via satellite, thanks to the GPRS network, sent to the DPC (Data Processing Center), located on the mainland. Any attempt to theft or tampering immediately triggers an alarm that alerts both the commander of the ship and the shipping company, thus minimizing the risk that something could happen to the cargo stowed on boats equipped with this system. Any possible abnormal condition detected is stored along with the date and time when it occurred. Additional information can be written on the tag memory, such as the name of the operator who has placed the seal, the cargo arrival date at the terminal and, of course, the number of the container.

Considering the import customs controls, despite a worsening of the timing of the activities in the port due to the need to re-initialize the seal at the container opening, a 85\% reduction in time is obtained to perform the "seen land" operation, while yet greater is the reduction in time for the operation of "seen exit" thanks to the automatic identification of the seal. This is true assuming that the containers arrive at the port already tagged: otherwise, if the Customs Agency provides for the tagging and the final recovery of the tags, the extra cost of this activity could make negative the investment.

Considering the operation of the terminal Information System, to be able to track the movements of each container during the periods of stay in the space of the port and the inner port, the RFID readers can be placed near the: (i) entrance to and exit from the storage area of the terminal of the port; (ii) entrance and exit from the storage area of the inner port manager; (iii) exit gate from the port; (iv) entrance and exit from the inner harbor.

The impact due to the RFID integration can be very relevant: (i) check on container number and integrity; (ii) check on the number and the seal integrity; (iii) counting of the containers entering or leaving the warehouse terminal; control at the gate for entry/exit; (iv) traceability of goods.

To date, the ports of Los Angeles (USA), Long Beach (USA), Hong Kong (China), Shenzhen, Chennai (India), Savannah (USA), Singapore, Rotterdam (Netherlands) and Jebel Ali implement the RFID technology [21, 22].

## 5.2 Remotely Piloted Aircraft

A promising wireless sensing technology is represented by drones, or "Remotely Piloted Aircraft" (RPA), aircrafts characterized by the absence of the human pilot on board. Today's drones can silently fly for more than 20 hours on the large area and collect information of all kinds, as well as being able to take chemical or radiological samples [25]. There are different models of drones, whose weight ranges from less than a kilogram to several tons.

Not carrying passengers, the drones are not pressurized and can fly at heights precluded to the airliners. The flight of drones that weigh several kilos is controlled by the computer on board the aircraft, under the remote control of a navigator or pilot on the ground or in another vehicle. The drones of greater weight are driven via satellite by complex earth stations. The advanced technology of these aircraft concerns the piloting system via satellite and anti-collision systems. A drone of the size of a light aircraft can stay airborne for 24 hours without refueling; for this reason, it is used for the surveillance of coasts and borders from which it continuously sends images to the ground station or other data collected from temperature or position sensors in order to prevent criminal activities (Figure 5.1). Small drones can be used for surveillance in ports, airports, nuclear power plants or other sensitive sites to prevent acts of terrorism or to check the goods from locations not accessible to man (Figure 5.2).

The drones, now consolidated by the extensive use for military purposes, are increasingly being used for civil applications, such as in operations for prevention and intervention in fire emergency, for use of non-military security, for surveillance of pipelines, with the purpose of remote sensing and research, often at a lower cost compared to conventional aircrafts [26]. In recent years, the technologies related to the development of the drones have undergone a rapid surge. In particular, the technological development in the field of cameras allows equipping a drone with multiple sensors, in the visible spectrum (compact or professional digital cameras) [27], infrared (thermal imaging), multi-spectral, till more advanced sensors, such as LIDAR [28] or sensors for monitoring the air quality.

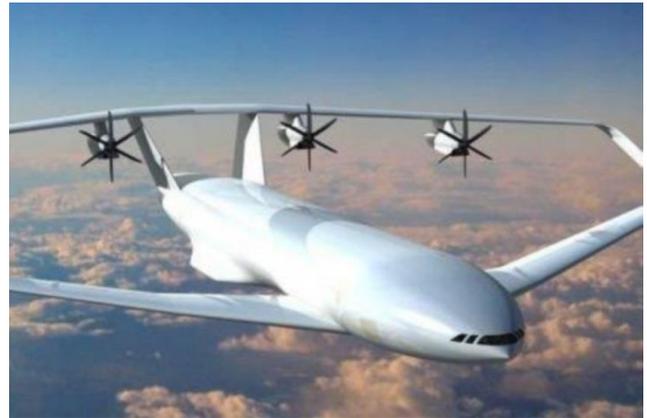

**Fig 5.1: Sample large drone.**

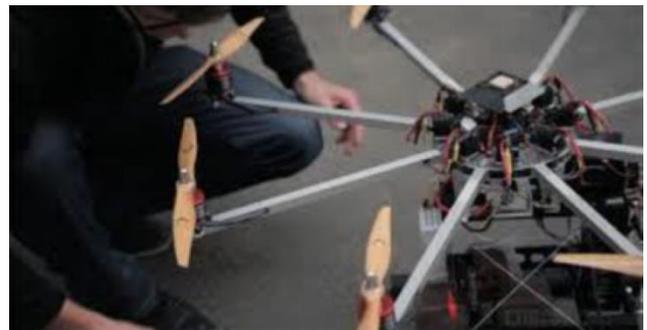

**Fig 5.2: Sample small drone.**

## 6. CONCLUSIONS

Port terminals are an example of complex logistics where the usage of ICT technologies has sensibly improved the management and the control of goods, also enhancing the logistics competitiveness. This paper presented an overview of the current available wireless technologies that can be used for data exchange and control in the port. In recognizing the pivotal role that ICT plays in modern custom administrations, an effort has to be done in proposing innovating ICT solutions with high priority for those technological solutions addressed to improve security, to monitor goods, to increase efficiency and time-spending.

## ACKNOWLEDGEMENT

This work has been co-funded by Italian MIUR (Ministry for Education, University and Research) in the framework of the PRIN project entitled "Eguaglianza Nei Diritti Fondamentali Nella Crisi Dello Stato e Delle Finanze Pubbliche: Una Proposta Per Un Nuovo Modello

Di Coesione Sociale Con Specifico Riguardo Alla Liberalizzazione e Regolazione Dei Trasporti", task "Sistema Di Monitoraggio A Distanza ed Automatizzato delle Merci e di Eventuali Persone Preposte Alla Sorveglianza Delle Stesse".

## REFERENCES

[1] Ehsan Ahvar and Mahmood Fathy. Special evaluation: A practical simulation-based method for accurate performance evaluation of routing protocols in wireless sensor networks. In Advanced Information Networking and Applications-Workshops, 2008. AINAW 2008. 22nd International Conference on, pages 1266–1271. IEEE, 2008.

[2] Wazir Zada Khan, NM Saad, and Mohammed Y Aalsalem. An overview of evaluation metrics for routing protocols in wireless sensor networks. In Intelligent and Advanced Systems (ICIAS), 2012 4th International Conference on, volume 2, pages 588–593. IEEE, 2012.

[3] Wang Xiaokai, Du Wenzhuo, and Zheng Dongmei. A class of the wireless sensor networks qos description and evaluation. In Measuring Technology and Mechatronics Automation (ICMTMA), 2011 Third International Conference on, volume 1, pages 16–19. IEEE, 2011.

[4] Yuan Lingyun and Wang Xingchao. Study on performance evaluation method based on measurement for wireless sensor network. In Communications Technology and Applications, 2009. ICCTA'09. IEEE International Conference on, pages 201–206. IEEE, 2009.

[5] Wang Jie, Kuan-jiu Zhou, Kai Cui, Xiang-jie Kong, and Guang Yang. Evaluation of the task communication performance in wireless sensor networks: a queue theory approach. In Green Computing and Communications (GreenCom), 2013 IEEE and Internet of Things (iThings/CPSCom), IEEE International Conference on and IEEE Cyber, Physical and Social Computing, pages 939–944. IEEE, 2013.

[6] KB Batra and Sanjeev Srivastava. Wireless sensor network-performance evaluation in the field. In Power, Control and Embedded Systems (ICPCES), 2012 2nd International Conference on, pages 1–8. IEEE, 2012.

[7] SIG Bluetooth. Specification of the bluetooth system, version 1.1. http://www.bluetooth.com, 2001.

[8] J Ben Othman, Hind Castel-Taleb, and Lynda Mokdad. Multi-services mac protocol for wireless networks. In Computer Systems and Applications, 2005. The 3rd ACS/IEEE International Conference on, page 61. IEEE, 2005.

[9] Matteo Bertocco, Giovanni Gamba, and Alessandro Sona. Is csma/ca really efficient against interference in a wireless control system? an experimental answer. In Emerging Technologies and Factory Automation, 2008. ETFA 2008. IEEE International Conference on, pages 885–892. IEEE, 2008.

[10] Ieee standard for information technology - telecommunications and information exchange between systems - local and metropolitan area networks - specific requirements - part 11: Wireless lan medium access control (mac) and physical layer (phy) specifications. IEEE Std 802.11-2007 (Revision of IEEE Std 802.11-1999), pages 1–1076, June 2007.

[11] Olayemi Olawumi, Keijo Haataja, Mikko Asikainen, Niko Vidgren, and Pekka Toivanen. Three practical attacks against zigbee security: Attack scenario definitions, practical experiments, countermeasures, and lessons learned. In Hybrid Intelligent Systems (HIS), 2014 14th International Conference on, pages 199–206. IEEE, 2014.

[12] Eko Nugroho, Alvin Sahroni, et al. Zigbee and wifi network interface on wireless sensor networks. In Electrical Engineering and Informatics (MICEEI), 2014 Makassar International Conference on, pages 54–58. IEEE, 2014.

[13] Keigo Nomura and Fumiaki Sato. A performance study of zigbee network under wi-fi interference. In Network-Based Information Systems (NBiS), 2014 17th International Conference on, pages 201–207. IEEE, 2014.

[14] RS Hsiao, DB Lin, HP Lin, CH Chung, and SC Cheng. Integrating zigbee lighting control into existing building automation systems. 2012.

[15] Yoshiharu Hirakata, Akira Nakamura, Kizuku Ohno, and Makoto Itami. Navigation system using zigbee wireless sensor network for parking. In ITS Telecommunications (ITST), 2012 12th International Conference on, pages 605–609. IEEE, 2012.

[16] Michele Girolami, Filippo Palumbo, Francesco Furfari, and Stefano Chessa. The integration of zigbee with the giraffplus robotic framework. In Evolving Ambient Intelligence, pages 86–101. Springer, 2013.

[17] Rashmi Bajaj, Samantha Lalinda Ranaweera, and Dharma P Agrawal. Gps: location tracking technology. Computer, 35(4):92–94, 2002.

[18] G Luchetti, G Servici, Emanuele Frontoni, Antonella Mancini, and Primo Zingaretti. Design and test of a precise mobile gps tracker. In Control & Automation (MED), 2013 21st Mediterranean Conference on, pages 1199–1207. IEEE, 2013.

[19] Mircea Popa and Bogdan Suta. A solution for tracking a fleet of vehicles. In Telecommunications Forum (TELFOR), 2011 19th, pages 1558–1561. IEEE, 2011.

[20] Qiaohong Zu and Haiyao Gao. Research on logistic information system based on rfid. In Computer-Aided Industrial Design and Conceptual Design, 2008. CAID/CD 2008.9th International Conference on, pages 409–412. IEEE, 2008.

[21] Mohamed K Watfa, Umaima Suleman, and Zaer Arafat. Rfid system implementation in jebel ali port. In Consumer Communications and Networking Conference (CCNC), 2013 IEEE, pages 950–955. IEEE, 2013.

[22] Shahriar Mohammadi and Sedigheh Ardast. A new model for evaluation of rfid utilization in various ports of the world. In Communications Technology and Applications, 2009. ICCTA'09. IEEE International Conference on, pages 36–39. IEEE, 2009.

[23] MIELE Project. http://www.onthemosway.eu/miele-2.

[24] Richard L Wilson. Ethical issues with use of drone aircraft. In Ethics in Science, Technology and Engineering, 2014 IEEE International Symposium on, pages 1–4. IEEE, 2014.

[25] Jet pilotless drones collect data on radiological hazards in atomic clouds. Electrical Engineering, 72(6):567–570, June 1953.

[26] Daniel Camara. Cavalry to the rescue: Drones fleet to help rescuers operations over disasters scenarios. In Antenna Measurements & Applications (CAMA), 2014 IEEE Conference on, pages 1–4. IEEE, 2014.

[27] Tiziana D'Orazio, Filippo Palumbo, and Cataldo Guaragnella. Archaeological trace extraction by a local directional active contour approach. Pattern Recognition, 45(9):3427–3438, 2012.

[28] Hans Karl Heidemann. Lidar base specification version 1.0.

[29] S Loviund, P Olsen, T Brokerud, and E Eilertsen. Comparative tests of dvb-s2&dvb-sin a full scale broadcasting system. In The Institution of Engineering and Technology Seminar on Digital Video Broadcasting Over Satellite: Present and Future, 2006.

[30] EN ETSI. 301 307 digital video broadcasting (dvb); v1. 1.2 (2006-06), second generation framing structure, channel coding and modulation systems for broadcasting, interactive services, news gathering and other broadband satellite applications, 2006. Available on ETSI web site (http://www. etsi. org).

[31] Mustafa Eroz, BF Beidas, Rohit Iyer Seshadri, and Lin-Nan Lee. Dvb-s2x: An update to dvb-s2. In Proc. 32nd AIAA Intern. Commun. Satellite Syst. Conference (ICSSC), 2014.

[32] Hui Hu and Chao Yuan. Performance analysis of galileo global position system. In Power Electronics and Intelligent Transportation System (PEITS), 2009 2ndInternational Conference on, volume 1, pages 396–399. IEEE, 2009.

[33] Jun-jie Peng. A survey of location based service for galileo system. In Computer Science and Computational Technology, 2008. ISCSCT'08. International Symposium on, volume 1, pages 737–741. IEEE, 2008.

[34] Mike Quinlan. Galileo-a european global satellite navigation system. In 2005 The IEE Seminar on New Developments and Opportunities in Global Navigation Satellite Systems (Ref. No. 2005/10810), 2005.

[35] Jin Cao, Maode Ma, Hui Li, Yueyu Zhang, and Zhenxing Luo. A survey on security aspects for lte and lte-a networks. Communications Surveys & Tutorials, IEEE, 16(1):283–302, 2014.

[36] Sadayuki Abeta. Toward lte commercial launch and future plan for lte enhancements (lte-advanced). In Communication Systems (ICCS), 2010 IEEE International Conference on, pages 146–150. IEEE, 2010.

[37] MFL Abdullah and AZ Yonis. Performance of lte release 8 and release 10 in wireless communications. In Cyber Security, Cyber Warfare and Digital Forensic (CyberSec), 2012 International Conference on, pages 236–241. IEEE, 2012.

[38] Ayman Elnashar, Mohamed El-Saidny, et al. Looking at lte in practice: A performance analysis of the lte system based on field test results. Vehicular Technology Magazine, IEEE, 8(3):81–92, 2013.

[39] Jan Oudelaar. Evolution towards umts. In Personal, Indoor and Mobile Radio Communications, 1994. Wireless Networks-Catching the Mobile Future., 5th IEEE International Symposium on, volume 3, pages 852–856. IEEE, 1994.

[40] PC Mason, JM Cullen, and NC Lobley. Umts architectures. 1996.

[41] M Coronado, Chandra S Lalwani, et al. Wireless vehicular networks to support road haulage and port operations in a multimodal logistics environment. In Service Operations, Logistics and Informatics, 2009. SOLI'09. IEEE/INFORMS International Conference on, pages 62–67. IEEE, 2009.

[42] Daan Pareit, Bart Lannoo, Ingrid Moerman, and Piet Demeester. The history of wimax: A complete survey of the evolution in certification and standardization for ieee 802.16 and wimax. Communications Surveys & Tutorials, IEEE, 14(4):1183–1211, 2012.

[43] Kejie Lu, Yi Qian, and Hsiao-Hwa Chen. Wireless broadband access: Wimax and beyond-a secure and service-oriented network control framework for wimax networks. Communications Magazine, IEEE, 45(5):124–130, 2007.

[44] Guifen Gu and Guili Peng. The survey of gsm wireless communication system. In Computer and Information Application (ICCIA), 2010 International Conference on, pages 121–124. IEEE, 2010.

[45] Friedhelm Hillebrand. The creation of standards for global mobile communication: Gsm and umts standardization from 1982 to 2000. Wireless Communications, IEEE, 20(5):24–33, 2013.